\documentclass[12pt]{iopart}

\usepackage{iopams}  

\usepackage[dvips]{color}

\usepackage{graphicx}

\def\mrm{\mathrm}

\def\goto{\rightarrow}

\def\mrm{\mathrm}

\def\goto{\rightarrow}

\def\vphi{\varphi}

\def\s0{\langle s_0 \rangle}
\def\nubar{\overline{\nu}}
\def\pbar{\overline{p}}

\begin{document}

\title[]
{Monte-Carlo simulation study 
of the two-stage percolation transition 
 in enhanced binary trees}

\author{Tomoaki Nogawa$^1$ and Takehisa Hasegawa$^2$}
\address{
$^1$ Department of Applied Physics, 
The University of Tokyo, 7-3-1 Hongo, Bunkyo-ku, Tokyo 113-8656, Japan
\\
$^2$ Department of Physics, Hokkaido University,
Kita-ku Kita 10-chome Nishi 8-chome, 
Sapporo, Hokkaido 060-0810, Japan}
\ead{nogawa@serow.t.u-tokyo.ac.jp}

\begin{abstract}
We perform Monte-Carlo simulations to study 
the Bernoulli ($p$) bond percolation 
on the enhanced binary tree which belongs to the class of 
nonamenable graphs with one end.
Our numerical results show that the system has 
two distinct percolation thresholds $p_{c1}$ and $p_{c2}$. 
The mean cluster size diverges as $p$ approaching 
to $p_{c1}$ from below. 
The system is critical at all the points in the intermediate phase 
$(p_{c1} < p  < p_{c2})$ 
and there exist infinitely many infinite clusters.
In this phase the corresponding fractal exponent 
continuously increases with $p$ from zero to unity. 
Above $p_{c2}$ the system has a unique infinite cluster. 
\end{abstract}

\pacs{64.60.ah, 68.35.Rh, 64.60.al, 89.75.Hc}
\maketitle

\section{Introduction}

\begin{figure}[t]
\begin{center}
\includegraphics[trim=160 418 300 -160,scale=0.4,clip]{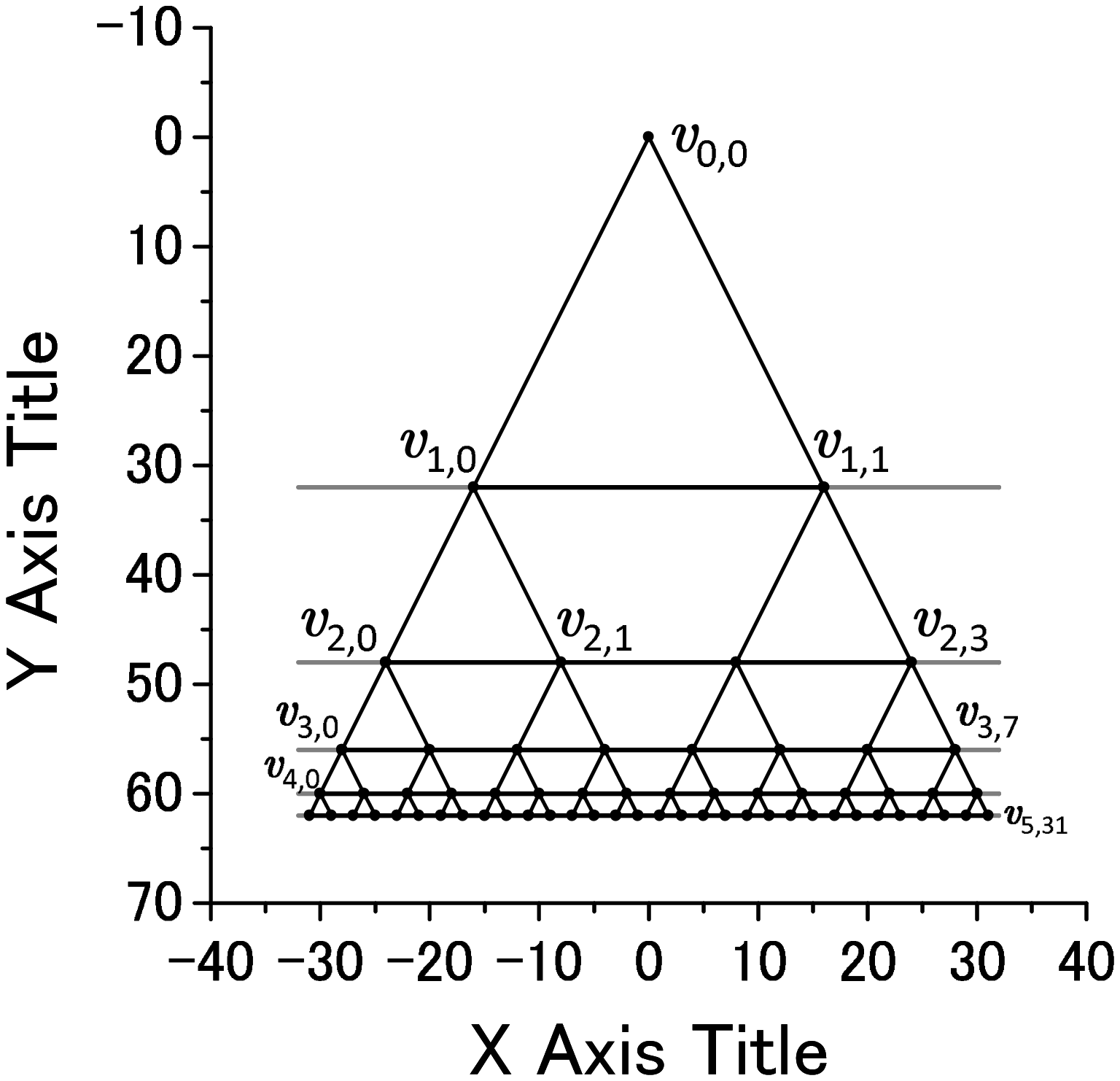}
\hspace{0.5cm}
\includegraphics[trim=70 348 150 -190,scale=0.3,clip]{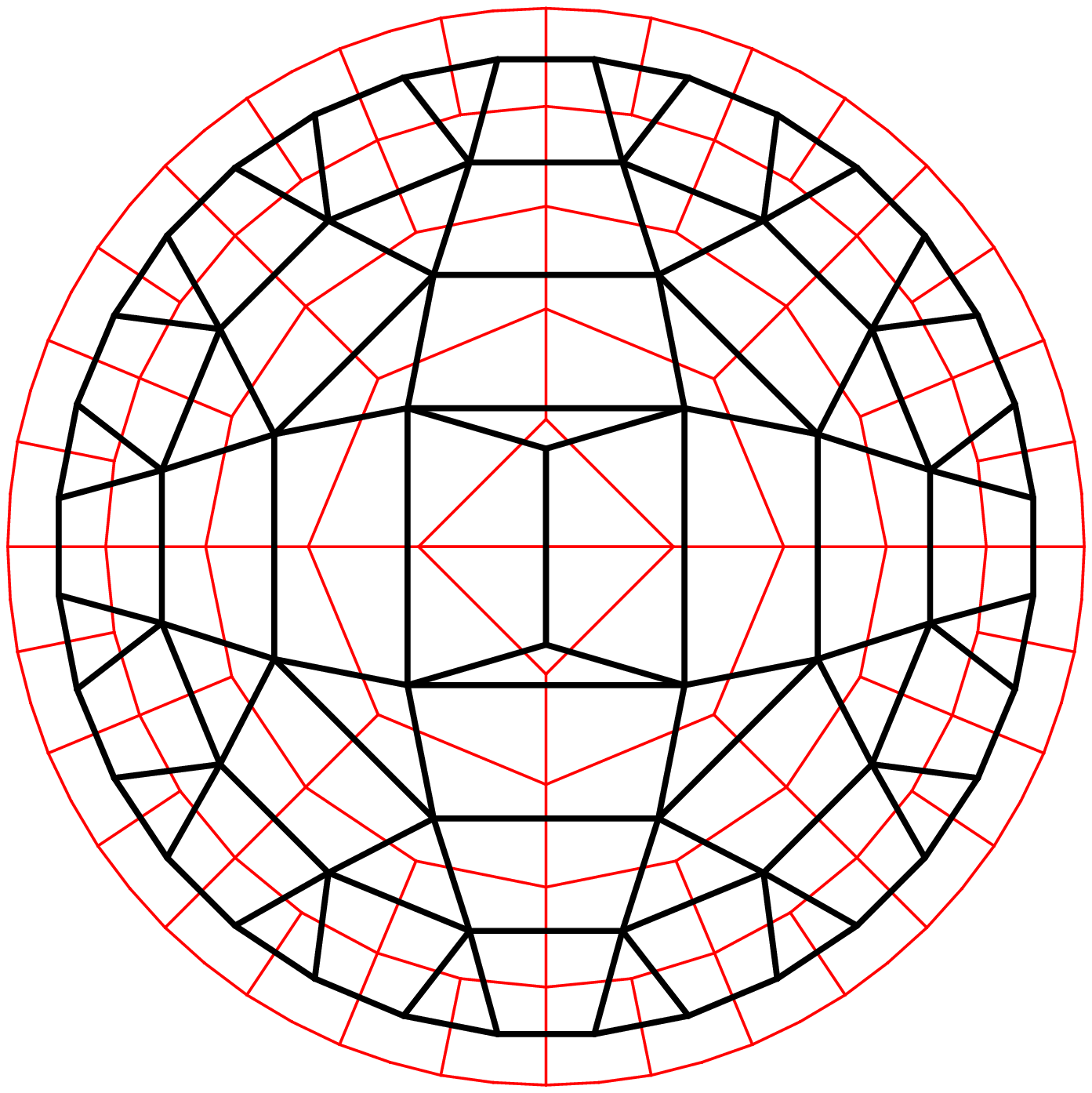}
\end{center}
\vspace{-3mm}
\caption{\label{fig:etb}
(left) 
Schematic diagram of the enhanced binary tree with 6 generations. 
The dangling bond on left and right side 
is connected in the periodic boundary condition.
(right) 
Another presentation of EBT (thick-black line) 
and its dual lattice (thin-red line). 
For easiness to see, the root vertex is not displayed for both the EBT and the dual lattice.
}
\end{figure}

Geometry of a space, which 
is characterized by dimensionality and topology, 
is a very important factor 
for the collective dynamics and statics of the elements embedded on it. 
It takes special significance in critical phenomena 
where correlation length diverges.   
Critical phenomena have been one of the main subjects 
of statistical physics and 
intensively studied over various systems 
such as interacting spin models, map cellular-automata, 
percolation, {\it etc}. 
In recent years, more and more attention has been paid to 
critical phenomena on
a class of graphs 
which have quite different properties 
with those on the well-studied Euclidean lattices. 
One representative example is the class, 
so-called 
blue{(almost)}
transitive 
{\it nonamenable graphs} (NAGs)  \cite{Lyons00, Schonmann01}, 
{\it e.g.}, Cayley trees, hyperbolic lattices  
and enhanced trees 
\cite{BS96} (see figure~\ref{fig:etb} for the binary case). 
Here ``transitive'' graph means that 
all vertices on it play exactly the same role 
in the large size limit.
The fundamental property of transitive NAGs is that 
the number of vertices in a ball centered at the origin 
increases exponentially with the radius. 
Recent studies have reported that 
this property crucially affects the collective dynamics 
to make it differ much from those on Euclidean lattices 
\cite{Lyons00, Schonmann01,Shima06a, Shima06b, Ueda07, Krcmar08,Wu96, Wu00, BS96, BS00}.
Although it seems that such graphs are non-realistic 
and studies on them are only for theoretical interests,
their distinctive features, so-called {\it small-world} property 
and a non-zero clustering coefficient, 
are common with {\it complex networks}, which are widely found in the social and natural systems \cite{Albert02, Newman03}.
For example, percolation problems 
on complex networks are related to 
error and attack tolerance and cascade failures of real world systems 
\cite{DorogoRev}. 
Instead of extensive studies 
the interplays between network structures and critical phenomena 
is still unclear.
Transitive NAGs has large advantages in its simplicity, 
e.g., regularity of the lattice, 
compared with usually studied complex networks.
Studies on them has have a possibility to give a breakthrough 
on this complex problem.

The number of {\it ends}, 
which indicates a kind of vulnerability of graphs, 
is an important concept for NAGs.
The number of ends $e(G)$ of an infinite graph $G$ is defined as 
the supremum of the number of infinite connected components in 
$G \backslash S$, where $G \backslash S$ is the graph obtained from $G$ 
by removing 
arbitrary finite subgraph $S$ and the bonds incident to those. 
While trees have infinitely many ends, 
hyperbolic lattices and enhanced trees have only one end. 
Recent analytical studies have indicated that 
the Bernoulli bond percolation on NAGs with one end exhibits
a new type of phase transition, so-called 
{\it multiple phase transition} (MPT),
which takes three distinct phases 
according to open bond probability $p$ 
as follows \cite{Lyons00, Schonmann01}: 
\begin{itemize}
\item{ {\it Nonpercolating phase}:  there is no infinite cluster for $0 \le p < p_{c1}(G)$.} 
\item{ {\it Intermediate phase}: there are infinitely many infinite clusters 
for $p_{c1}(G)< p <p_{c2}(G)$.} 
\item{ {\it Percolating phase}: there is a unique infinite cluster for $p_{c2}(G) < p \le 1$.}
\end{itemize}
Here {\it infinite cluster} means 
a cluster whose mass diverges with system size $N$ 
as $N^\psi$ with $0<\psi \le 1$. 
As well known for the Euclidean lattices, 
any amenable graph, {\it e.g.}, can have at most only one infinite cluster, 
i.e., $p_{c1}=p_{c2}$ \cite{BK89}. 
By contrast there are always plural infinite clusters in the percolating phases of trees. 
This is a consequence of infinitely many ends, i.e., 
trees have no long cyclic paths and split easily.
Both of an exponential growth and sufficient amount of loops are 
necessary for graphs to exhibit three distinct phases. 
Indeed, Benjamini and Schramm proved the existence of a MPT 
on any planar transitive NAG with one end \cite{BS00}. 
In addition, it is proved that the transition 
at $p_{c1}$ belongs to the mean-field universality class \cite{S02}.

Previous studies of MPTs have been restricted in the analytic way 
based on probability theory and we lack numerical research, 
which will bring us the quantitative indication of MPTs. 
Especially finite size effect 
is hardly investigated by the analytic approach. 
This effect is crrucial 
to argue the NAGs and the realistic network, 
where the typical length scale of the graphs 
often grows as a logarithmic function of 
the total number of vertices $N$ 
and is still small for considerably large $N$. 
In this paper, we investigate a MPT by Monte-Carlo simulations 
taking the enhanced binary tree (EBT) as an example. 
Our numerical results actually give two distinct critical probabilities 
$p_{c1}$ and $p_{c2}$. 
All the points in the intermediate phase $(p_{c1} < p  < p_{c2})$ 
are critical 
and there exist infinitely many infinite clusters .
The corresponding fractal exponent 
continuously increases with $p$ from 0 to 1 
and bridges a gap between non-percolating and percolating phases.

\section{Fundermental mechanism of the two-stage transition}
\label{sec:picture}

In this section we provide a basic picture of MPTs on NAGs. 
Although we consider the EBT as an example, 
the concept is not specified to it. 

At first we define an enhanced binary tree with $L$ generations. 
We prepare vertices $v_{n,m}$ for 
$n = 0, 1, \cdots, L-1$ and $m = 0, 1, \cdots,  2^n -1$. 
The total number of vertices $N$ equals $2^L-1$, 
i.e., $L \approx \log_2 N$ . 
Inter-generation (radial) bonds are supposed 
between $v_{n,m}$ and $v_{n+1,2m}$ and 
between $v_{n,m}$ and $v_{n+1,2m+1}$.
These bonds yield a binary tree. 
Furthermore intra-generation (circumferential) bonds are added 
between $v_{n,m}$ and $v_{n,m+1}$, 
which means ``enhanced''. 
All of inter- and intra-generation bonds are open 
with homogeneous probability $p$ 
to yield connected clusters. 
EBT is not exactly {\it transitive} but {\it almost} transitive 
since the root vertex $v_{0,0}$ is special 
and the center can be defined.
But this presumably does not affect 
the essential property of the critical phenomena.

The following scenario is based on the natural assumption that 
the connectedness (correlation) function; 
the probability that a vertex of the $\ell$-th generation 
is connected to the root clster, 
that $v_{0,0}$ belongs to, exponentially decays wiht $\ell$ as 
\begin{equation}
C_0(\ell) \propto 2^{-\ell/\xi}.
\label{eq:c0}
\end{equation}
Here $\xi$ means a correlation length 
along the generation (radial) direction 
and it monotonically increases with 
bond opening probability $p$. 
The averaged mass of the root cluster $\s0$ is calculated 
by summing up the connectedness function as 
\begin{equation}
\s0 \propto \sum_{\ell=0}^{L-1} 2^\ell C_0(\ell) 
= \frac{\alpha^L - 1}{\alpha-1} 
\quad \mrm{where} \quad
\alpha \equiv 2^{1 - 1/\xi} 
\label{eq:mass}
\end{equation}
and $2^\ell$ is the number of vertices at the $\ell$-th generation 
taking a role of surface area factor. 
For $\xi<1$ and $\alpha<1$, 
$\s0$ converges to finite value 
in the thermodynamic limit $L \rightarrow \infty$. 
On the other hand $\s0$ diverges with $L$ 
for $\xi \ge 1$ and $\alpha \ge 1$. 
It is important that the mass of cluster diverges 
despite of finite correlation length 
owing to the exponentially diverging surface area. 
This is an exact mechanism for the percolation transition 
on the binary Cayley tree, where 
$C_0(\ell)=p^\ell$ and $\xi=1/\log_2 (1/p)$. 
We can define the critical probability $p_{c1}$ 
at which a correlation {\it mass} diverges. 
From equation (\ref{eq:mass}), $\s0$ diverges very slowly 
with respect to the system size as 
\begin{equation}
\s0_L \propto L \approx \log_2 N \quad \mrm{at} \quad p=p_{c1}.
\end{equation}
Above $p_{c1}$, $\s0$ diverges with the system size as 
\begin{equation}
\s0_L \propto \alpha^L = 2^{L \log_2 \alpha} = N^\psi.
\end{equation}
where
\begin{equation}
\psi \equiv \log_2 \alpha = 1 - 1/\xi.
\end{equation}
The exponent $\psi$ mimics $d_f/d$ 
where $d_f$ is the fractal dimension 
of a cluster embedded on the $d$-dimensional Euclidean lattice. 
Note that $\psi$ is a function of $\xi$ and increases with $p$. 
Even though a cluster mass diverges above $p_{c1}$, 
the ratio $\s0/N$, which can be regarded as an order parameter,  
goes to zero for $N \rightarrow \infty$ as far as $\psi < 1$. 
If $\xi$ diverges at a certain $p$, 
$\psi$ continuously approaches to 1. 
We define the critical probability $p_{c2}$ at which 
the correlation {\it length} diverges. 
At this critical point a prefactor of the connectedness function 
in equation (\ref{eq:c0}) 
in form $\ell^{-\tilde{\eta}}$ takes an important role. 
The exponent $\tilde{\eta}$ corresponds to 
$2-d+\eta$ for the $d$-dimensional Euclidean lattice.
Positive $\tilde{\eta}$ makes $\s0/N$ vanish in the thermodynamic limit 
and the order parameter continuously rising from zero at $p_{c2}$. 
In this case, however, 
$\s0/N$ at $p_{c2}$ approaches to zero very slowly with $N$ 
as $(\log_2 N)^{-\tilde{\eta}}$.


The finite size dependences are summarized as 
\begin{equation}
\s0 \propto
\left \{
\begin{array}{ccl}
N^0 & \mrm{for} & p<p_{c1} \\
\log_2 N & \mrm{at} & p=p_{c1} \\
N^{\psi(p)} & \mrm{for} & p_{c1}<p<p_{c2} \\
N/(\log_2 N)^{-\tilde{\eta}}  & \mrm{at} & p = p_{c2}  \\
N & \mrm{for} & p > p_{c2}.
\end{array}
\right.
\end{equation}
The cluster mass shows fractal behavior in the phase $p_{c1}<p<p_{c2}$ 
in contrast with the Euclidean lattice 
which has a critical {\it point} and unique fractal exponent.
We consider the discrepancy of the probabilities 
for correlation mass and length 
is the key concept of the MPT on NAGs, 
which never occurs on the Euclidean lattices. 
In the following section we confirm the validity 
of the scenario shown above by numerical simulations.

\section{Numerical results}

Here let us explain 
the detail of our Monte-Carlo simulations. 
The systems with $L=10-22$ are investigated.
We generate samples, i.e., 
sets of open and closed bonds, 
by using pseudo random number generated 
by the Mersenne-Twister method \cite{MN98}. 
The ensemble average of observed quantities 
is taken over 480,000 samples. 
In order to improve the precision of the average values for small $p$, 
we treat the bonds which is connected to 
initial 3 generations exactly, 
i.e., counting all $2^{3 \times (2^3-1)-2}$ possible realizations 
with probability $p^{n_o} (1-p)^{n_c}$, 
where $n_o$ ($n_c$) is the number of open (closed) bonds. 
This treatment only takes the CPU-time independent of $N$.

We also investigate the percolation on the dual lattice of the EBT 
(see figure~\ref{fig:etb}), 
which is also a NAG with one end. 
Each vertex of the dual lattice is put 
on the center of triangle or rectangle cells of the EBT 
and each bond, 
which is open with probability $\pbar$, 
crosses with the conjugate bond of the EBT. 
A duality relation 
\begin{eqnarray}
p_{c1} + \overline{p}_{c2} = 1 
\quad \mrm{and} \quad
p_{c2} + \overline{p}_{c1} = 1
\label{eq:duality}
\end{eqnarray}
is known for dual planar NAGs \cite{BS00}.

\begin{figure}
\begin{center}
\includegraphics[trim=30 240 20 -225,scale=0.35,clip]{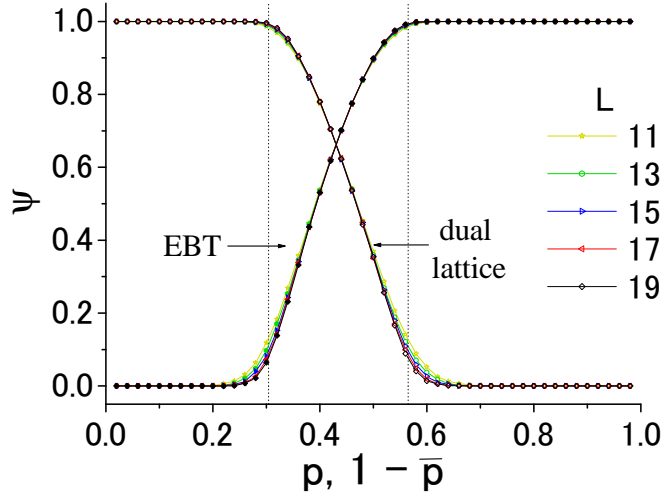}
\end{center}
\vspace{-3mm}
\caption{\label{fig:psi-p}
The $p$-dependence of $\psi$ for the enhanced binary tree 
and its dual lattice. 
For the latter, the horizontal axis is $1-\pbar$ 
to cofirm the duality relation. 
Results for various $L$'s are shown together.
The two vertical lines indicates $p=0.304$ and $p=0.563$, 
respectively.
}
\end{figure}

As mentioned in the previous section
the $N$-dependence of $\s0$ for the finite size systems 
determines the expected three phases. 
So we estimate a fractal exponent 
\begin{equation}
\psi = \frac{d \ln \s0_N}{d \ln N}
\approx \frac{d}{d L} \log_2 \s0_N, 
\label{eq:psi_def}
\end{equation}
so that $\s0 \propto N^\psi$. 
In practice, we calculate $\psi$ by the difference 
$[\ln \s0_{2N}-\ln \s0_{N/2})]/[\ln(2N)-\ln(N/2)]$.
This is a good approximation of equation~(\ref{eq:psi_def}) 
for large $N$ when $\s0$ is a power function of $N$.
The $p$-dependences of $\psi$ 
for both of the EBT and the dual lattice 
are shown in figure~\ref{fig:psi-p}. 
Three distinct phases can be observed as expected. 
The exponent $\psi$ of the EBT grows from zero to unity 
in the intermediate phase 
between $p \approx 0.30$ and $p \approx 0.56$. 
Oppositely $\psi$ of the duality lattice 
decrease from unity to zero 
between $1-\overline{p} \approx 0.30$ and $1-\overline{p} \approx 0.56$.
This suggests the duality relation actually holds.

\begin{figure}
\begin{center}
\includegraphics[trim=30 240 180 -225,scale=0.35,clip]{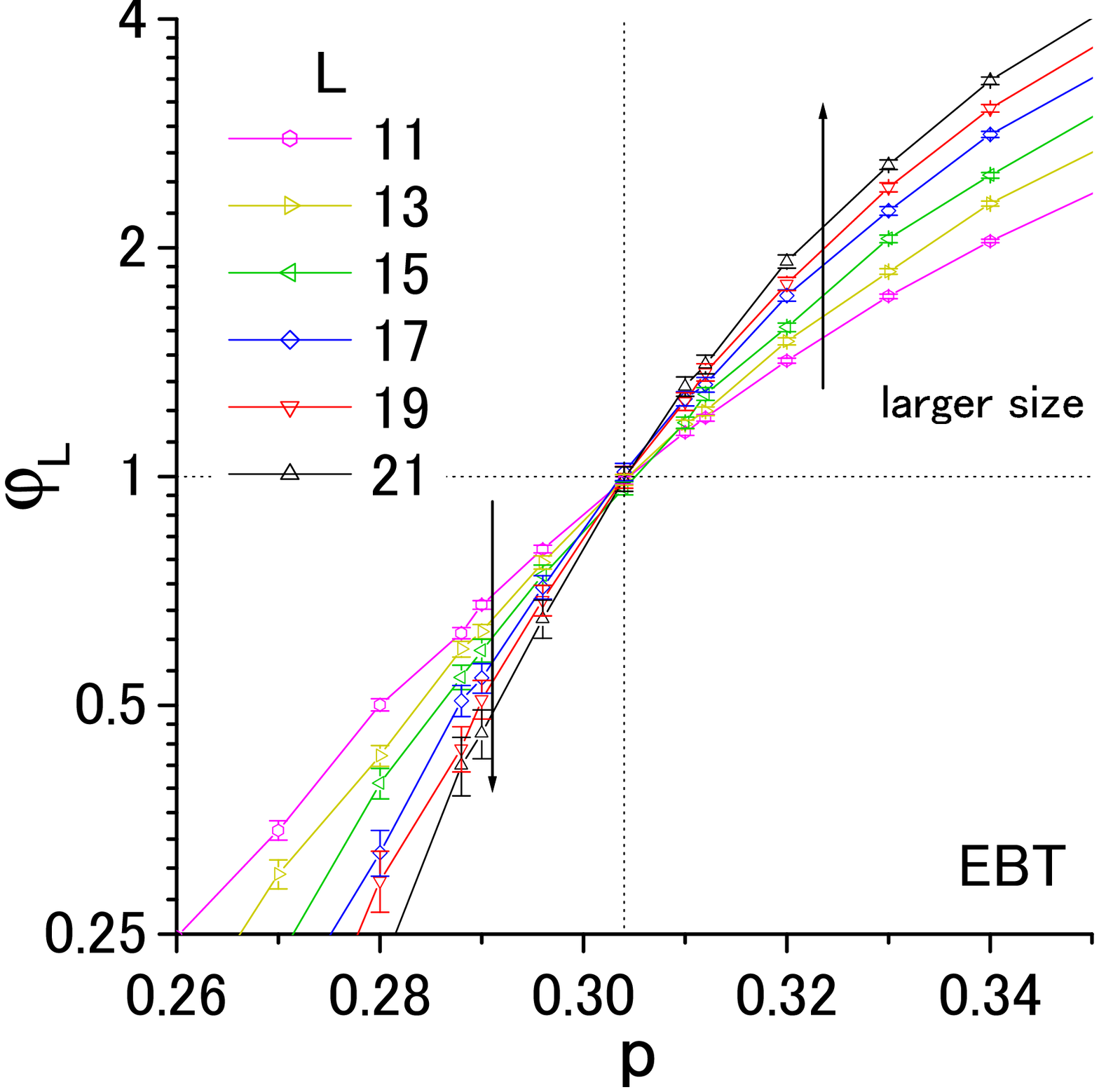}
\includegraphics[trim=30 240 180 -225,scale=0.35,clip]{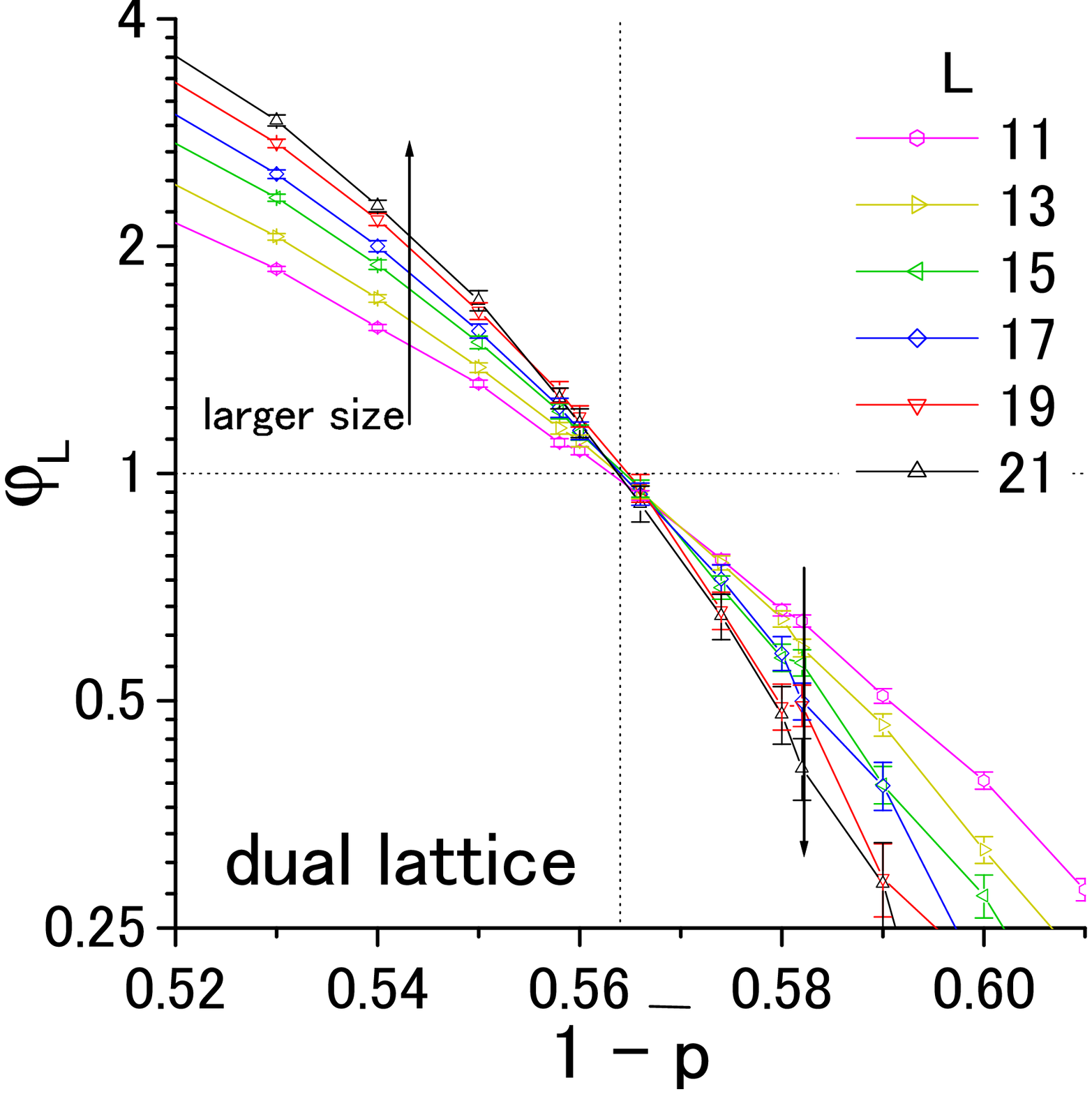}\end{center}
\vspace{-3mm}
\caption{\label{fig:phi-p}
The $p$-dependence of $\vphi_L$ 
for the EBT (left) and its dual lattice (right) with several system sizes. 
The crossing point indicates the transition point 
where $\langle s_0 \rangle$ logarithmically diverges. 
}
\end{figure}

The $N$-dependence of $\s0$ at $p_{c1}$ 
is decided as follows.
Approaching $p_{c1}$ from above, one expects that 
$\psi$ goes to zero 
but $\s0$ diverges as $\ln N$ ($\propto L$) at $p=p_{c1}$. 
To exclude the possibility that $\s0 \propto L^{\vphi}$ 
with $\vphi \ne 1$, we plot 
\begin{equation}
\vphi_L(p) = \frac{d \ln \s0_N}{d \ln L}, 
\end{equation}
as a function of $p$ in figure~\ref{fig:phi-p}. 
As $L$ goes to infinity, 
$\vphi_L(p)$ decreases to zero for $p<p_{c1}$ 
and diverges for $p>p_{c1}$. 
Only at $p=p_{c1}$, $\vphi_L$ converges 
to $N$-independent value, 
which is estimated at unity. 
Practically we obtain the precise critical probability 
$p_{c1}=0.304(1)$ from this crossing point 
and $\overline{p}_{c1}=0.436(1)$ in the same way.
Using the duality relation (\ref{eq:duality}), 
$p_{c2}=0.564(1)$ and $\overline{p}_{c1}=0.696(1)$ 
are also determined.

\begin{figure}[t]
\begin{center}
\includegraphics[trim=217 302 160 -165,scale=0.25,clip]{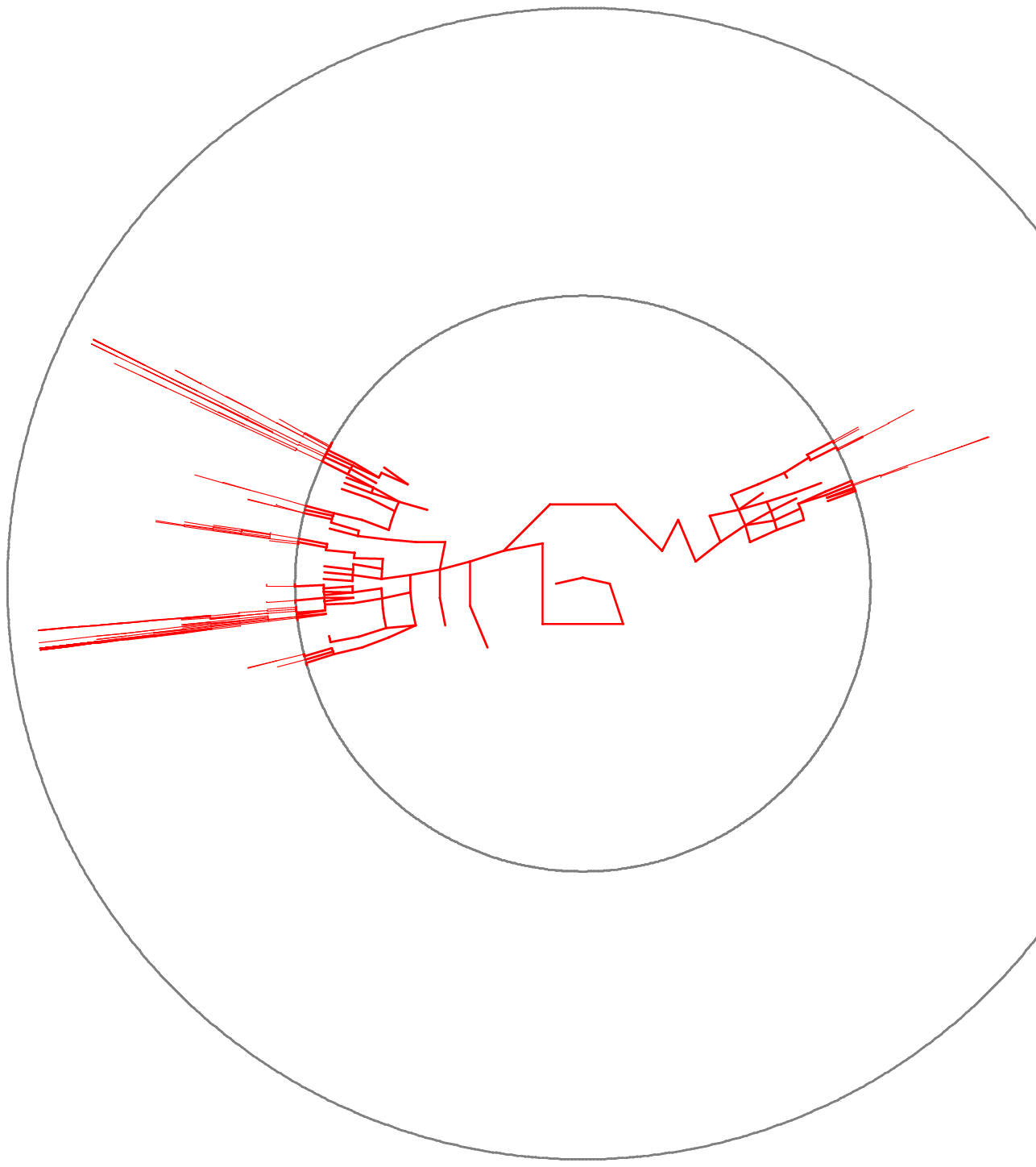}
\includegraphics[trim=217 302 160 -165,scale=0.25,clip]{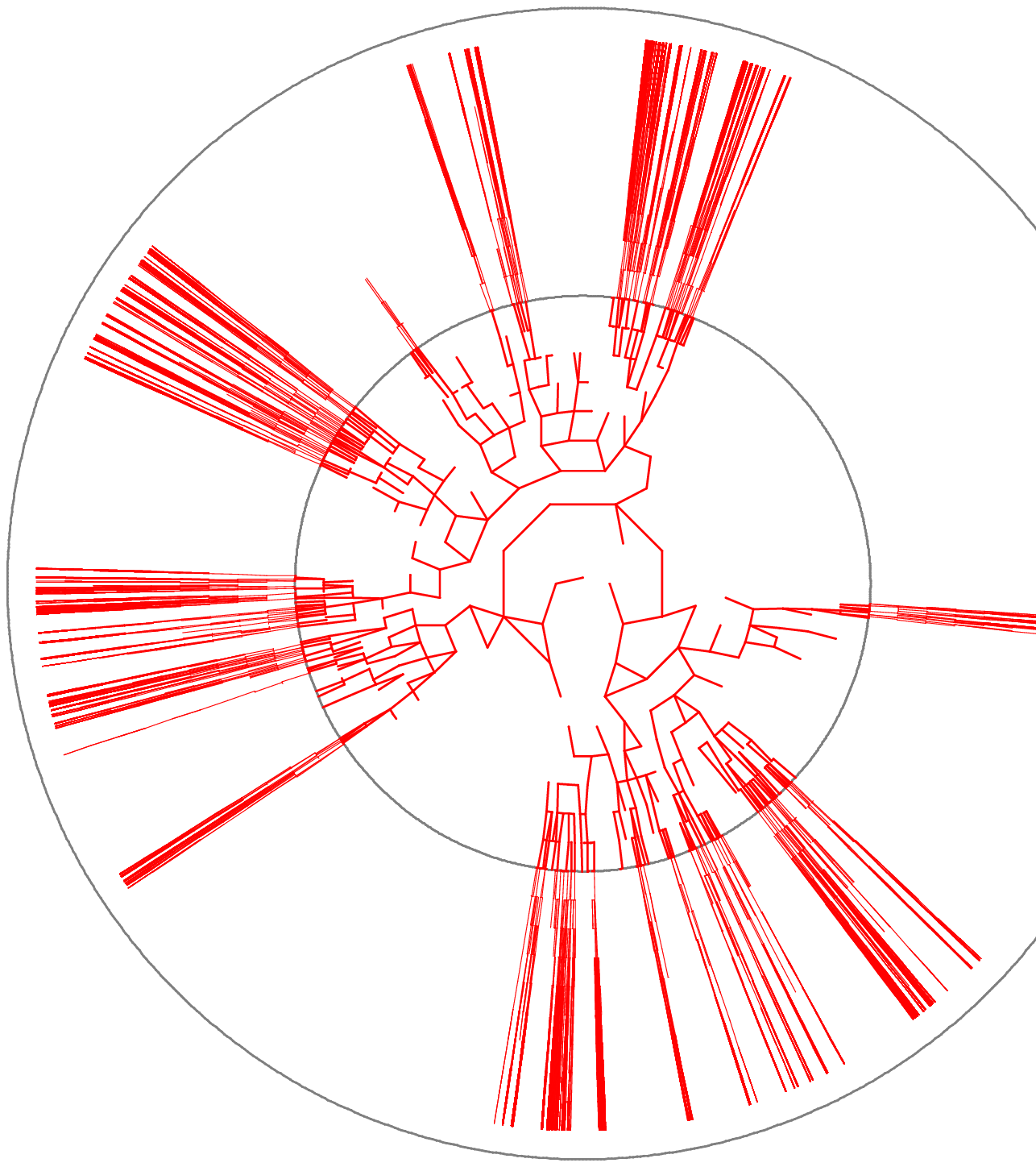}
\includegraphics[trim=0 0 0 0,scale=0.0605,clip]{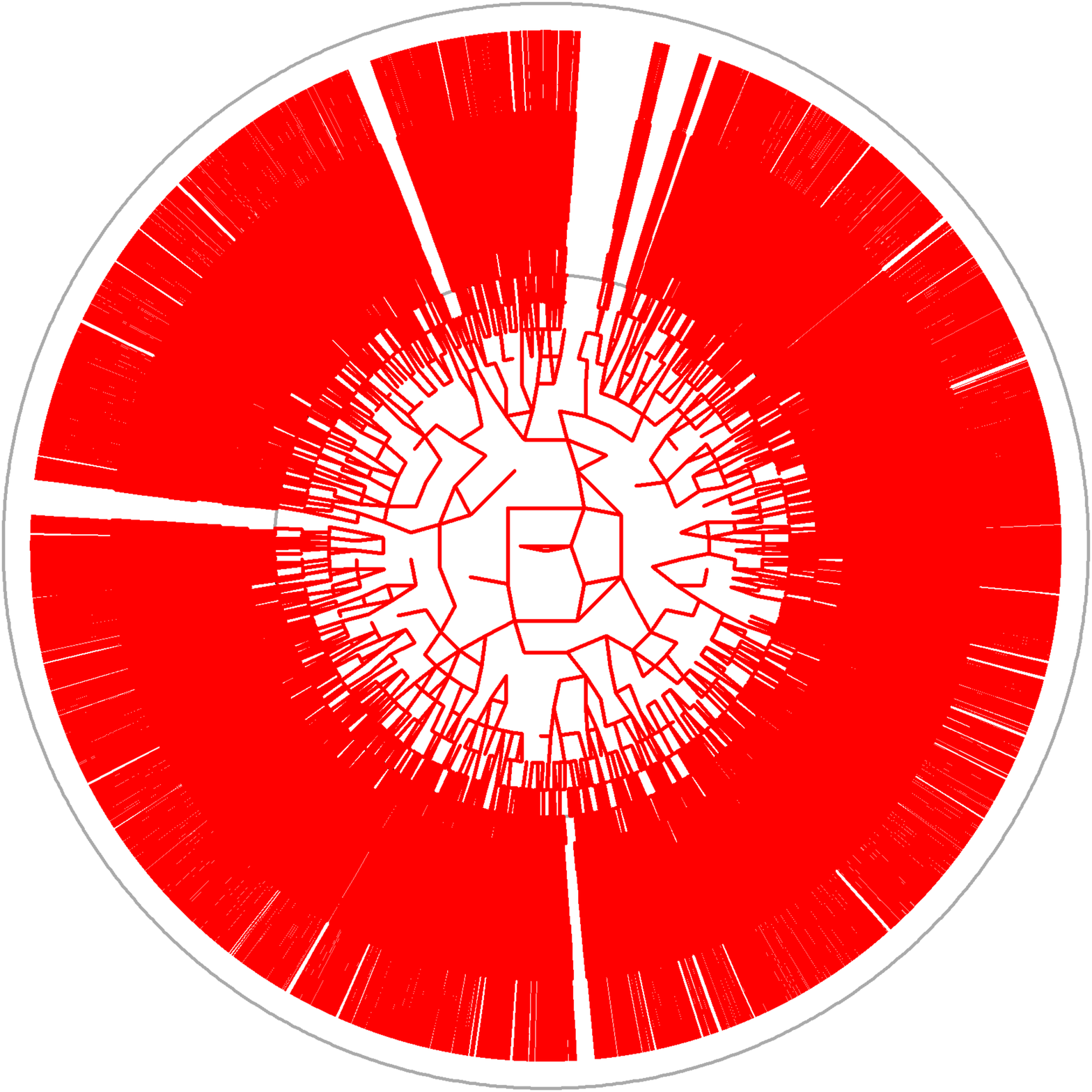}
\\
(a)\hspace{2.4cm}(b)\hspace{2.4cm}(c)\\
\end{center}
\vspace{-3mm}
\caption{\label{fig:cl_graph}
Typical geometry of giant clusters in the system with $L=20$ 
for $p$=0.304(a), 0.400(b), 0.564(c). 
}
\end{figure}

In figure \ref{fig:cl_graph} 
we show the geometry of the root cluster on the EBT 
by choosing samples which has a relatively large size cluster. 
Figure \ref{fig:cl_graph}(a) is for $p=0.304 \approx p_{c1}$. 
The cluster survive marginally; its branches do not show spreading behavior,  
and therefore the mass of the cluster is proportional 
to the number of generations. 
Figure \ref{fig:cl_graph}(b) shows the cluster 
at a middle point in the intermediate phase, which has many branches. 
The branches are spreading with the generation and tend to diverge. 
However this cluster occupies very small part 
of the whole system because spreading rate is slower than 
that of the EBT itself. 
Consequently, there is a room for other infinitely large clusters. 
Figure \ref{fig:cl_graph}(c) shows the cluster at $p=0.564\approx p_{c2}$. 
It looks compact and will occupy the finite fraction of 
the whole system in the thermodynamic limit.

\begin{figure}
\begin{center}
\includegraphics[trim=30 240 20 -225,scale=0.35,clip]{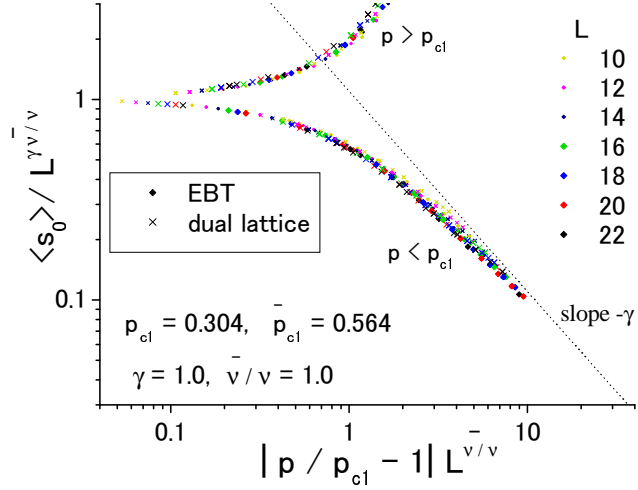}
\end{center}
\vspace{-3mm}
\caption{\label{fig:s0-p}
Finite size scaling plot of $\langle s_0 \rangle$ 
as a function of $p-p_{c1}$. 
The two groups, the EBT and the dual lattice, are plotted 
in arbitrary unit to collapse. 
}
\end{figure}

Next we estimate the critical exponents of the first transition at $p=p_{c1}$.
Let us suppose a finite size scaling law 
\begin{equation}
\s0 \propto L^{\gamma \nubar / \nu} 
\tilde{s}_0 \left( ( p_{c1} - p ) L^{\nubar / \nu} \right), 
\end{equation}
for $p<p_{c1}$. 
\footnote{
Consult Ref.~\cite{Havlin} about the exponent $\nubar$. 
We use generation $L$ as a characteristic length 
instead of {\it chemical distance} in these papers.
The number of retrenched horizontal paths is so small 
compared with vertical ones 
that we regard this change to be irrelevant.
}
The scaling function $\tilde{s}_0(\cdot)$ should 
have asymptotic forms, 
\begin{equation}
\tilde{s}_0(x) \propto
{\Biggl\{}
\begin{array}{ccl}
\mrm{const.} & \mrm{for} & x \ll 1 \\
x^{-\gamma} & \mrm{for} & x \gg 1. \\
\end{array}
\end{equation}
Figure~\ref{fig:s0-p} shows a nice collapsing of data 
by using $\gamma=1.0$ and $\nubar/\nu=1.0$. 
The data for the dual lattice also shows same scaling behavior 
(the form of scaling function is very similar to the one for the EBT). 
This result confirms
that the first transition at $p_{c1}$ 
belongs to the mean field universality class 
($\gamma=1$ and $\nubar=\nu=1/2$ \cite{Havlin}) 
as predicted in Ref.~\cite{S02}.
On the other hand the transition at $p_{c2}$ does not look 
like an usual continuous phase transition. 
The order parameter $\s0/N$ is fitted well by 
$\s0/N = 0.49+0.58L^{-0.083}$ at $p=0.564\approx p_{c2}$ 
using data for $L=10-24$ (not shown here). 
This means $\tilde{\eta}=0$ 
$\s0/N$ has a finite limit value in $L \goto \infty$ 
at the critical point. 
We have to be careful to conclude such special feature, 
discontinuous transition in the framework of second order transition, 
but can say, at least, that $\tilde{\eta}$ is smaller than 0.083.

\begin{figure}
\begin{center}
\includegraphics[trim=30 240 20 -225,scale=0.35,clip]{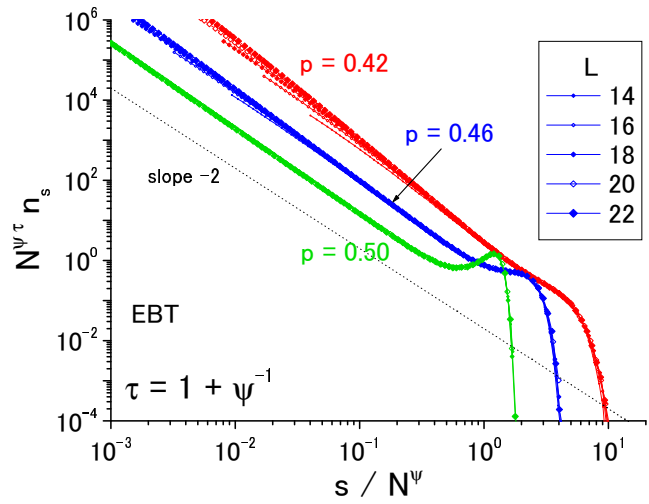}
\end{center}
\vspace{-3mm}
\caption{\label{fig:n-s}
Finite size scaling plot of $n_s$ as a function of $s$ 
for several $p$'s in the intermediate phase. 
We use the values of $\psi$ shown in figure~\ref{fig:psi-p}.
Here we eliminate the data for ($s<16$) since 
the data with too small $s$ does not obey to the scaling law. 
}
\end{figure}

Finally we investigate the distribution function $n_s$, 
which is the number of clusters with size $s$ per vertex, 
in order to show that the system is always critical 
in the intermediate phase 
as already implied from the $N$-dependence of $\s0$. 
We assume a finite size scaling law  
\begin{eqnarray}
n_s(N) = N^{-\psi \tau} \tilde{n}( s  N^{-\psi} ), 
\label{eq:ns-scaling}
\end{eqnarray}
for fixed $p$ between $p_{c1}$ and $p_{c2}$. 
Here the scaling function $\tilde{n}(x)$ is 
a power function $x^{-\tau}$ for $x \ll 1$, 
and a rapidly decreasing function for $x \gg 1$. 
In the scaling plot, figure~\ref{fig:n-s}, 
we use $\psi$ evaluated by equation (\ref{eq:psi_def}) 
and the scaling relation $\tau = 1+\psi^{-1}$ at each $p$
(note that this relation is based on the scaling relation 
$\tau = 1 + d_f/d$ on the Euclidean lattices).
Figure~\ref{fig:n-s} strongly supports that 
the above finite size scaling (\ref{eq:ns-scaling}) 
holds in the intermediate phase. 
Equation~(\ref{eq:ns-scaling}) indicates that 
the characteristic size of clusters diverges with $N \rightarrow \infty$ 
and a power-law distribution, $n_s(s) \propto s^{-\tau}$, holds 
up to infinite $s$. 
When such a power-law exists, 
it can be said that 
the number of infinite clusters is infinite 
because one can always find $\psi'$ satisfying $0<\psi'<\psi$ 
for any positive $\psi$ with which 
the number 
\begin{eqnarray}
N(s>N^{\psi'}) \equiv 
\int_{N^{\psi'}}^\infty ds N n_s(N)  
\sim N^{(1-\psi'/\psi)}, 
\end{eqnarray}
diverges in the limit $N \rightarrow \infty$.

\section{Summary and discussions}

In conclusion, 
we studied the bond percolation problem 
on the enhanced binary tree 
by Monte Carlo simulations for the first time. 
We observed the intermediate phase 
in addition to the nonpercolating and percolating phases 
where the average cluster size is finite 
and in the same order of the whole system size $N$, respectively.
All the points in the intermediate phase is critical 
and the fractal exponent $\psi$, where $\s0 \propto N^\psi$, 
increases continuously from zero to unity with increasing $p$.
Note that the system looks critical in terms of mass of cluster, 
and the correlation length must diverge at $p_{c2}$.

We expect that the mechanism of the MPT on the EBT 
is commonly observed in the other systems of NAGs with one end. 
There is, however, an interesting open problem, 
for both analytic and numerical studies, 
whether non-transitivity or inhomogeneity 
changes the scenario or not,  
which seriously corresponds to the complex networks.

\ack

The present work is supported by 21st Century COE program 
``Topological Science and Technology''.
We wish to thank H.Shima and Y.Sakaniwa for their stimulating works, 
and K.Nemoto and K.Yakubo for helpful comments and discussions.

\end{document}